# III-V semiconductor nano-resonators—a new strategy for passive, active, and nonlinear all-dielectric metamaterials

*Sheng Liu,* [1,2,*] *Gordon A. Keeler*[1], *John L. Reno*[1,2], *Michael B. Sinclair*[1], *Igal Brener*[1,2,*]

[1]Sandia National Laboratories, Albuquerque, NM 87185, USA
[2]Center for Integrated Nanotechnologies, Sandia National Laboratories, Albuquerque, NM 87185, USA
E-mail: snliu@sandia.gov, ibrener@sandia.gov

Metamaterials provide the freedom to engineer electromagnetic properties and behaviors that go beyond what is found in natural materials. Despite the success of metallic metamaterials at radio-frequencies, the intrinsic ohmic loss of metals at optical wavelengths limits their performance for practical applications. All-dielectric metamaterials that exploit the Mie resonances of high-permittivity resonators are a promising alternative for low loss operation at optical frequencies. Dielectric metamaterials were initially realized at radio and microwave frequencies,[1] where it was shown that arbitrary permittivities and permeabilities can be obtained through excitation of the tightly confined Mie resonances. Recently, there has been a flurry of activity surrounding all-dielectric metasurfaces at optical frequencies such as: metasurfaces made from Tellurium-based cubic resonators operating in the thermal infrared (IR);[2-4] and a number of epitaxial or amorphous silicon metasurfaces operating in the visible and near-IR.[5] The latter have shown a number of interesting properties and have enabled applications such as high efficiency Huygens' metasurfaces,[6] zero-index emission,[7] beam steering,[8] vortex beam generation,[9] high efficiency third harmonic generation,[10, 11] and many other ultrathin optical components.[12-14]

Using Si to scale high-performance all-dielectric metasurfaces to visible wavelengths[15-17] is challenging since it absorbs wavelengths above its bandgap (~1 μm). Moreover, due to the indirect bandgap of Si and its poor light emission, integration of Si-based dielectric metasurfaces and metamaterials with light emitting structures is challenging. In contrast, metasurfaces and metamaterials made from direct bandgap III-V semiconductors could pave the way to new fundamental physics and applications combining emitters, detectors, and nonlinear optical behavior. The use of III-V semiconductors could also mitigate other drawbacks of Si that could affect optical switching applications, including strong free carrier absorption due to the long electron lifetime in indirect bandgap materials. For emerging studies of nonlinear optical metamaterials, III-V semiconductors offer very high second-order susceptibilities that are absent in Si due to its centrosymmetric crystal structure.[10, 18]

In this paper, we will demonstrate GaAs metasurfaces fabricated using a combination of high aspect ratio etching and selective wet oxidation of AlGaAs under-layers to form a low refractive index oxide. Our new approach avoids complex flip-chip bonding processes[19] that have previously been used for GaAs resonator arrays. The experimental reflectivity spectra of our GaAs resonator arrays exhibit high reflectivity peaks at electric and magnetic dipole resonances that shift as expected as the resonator size varies. We further use the same fabrication processes to demonstrate quasi-3D GaAs dielectric resonator arrays that provide us with new degrees of freedom in device engineering. For these arrays, we experimentally measure ~100% reflectivity over a broad spectral range. Finally, we numerically show two examples of new functionality and applications that could be enabled by these multilayer dielectric metamaterials.

High refractive index contrast between dielectric resonators and their surrounding media is essential for good confinement of the electromagnetic fields inside the resonators — a necessary condition for realizing most functions using dielectric metamaterials. However, different alloys of epitaxially grown III-V semiconductors have similar refractive indices, and thus far, high index contrast has only been achieved using complex flip-chip techniques that bond the resonators to lower index substrates.[19] Here, we create (Al)GaAs based dielectric metasurfaces by selectively oxidizing AlGaAs to form $(Al_xGa_{1-x})_2O_3$. The refractive index of this native oxide is ~1.6, which is sufficiently lower than GaAs[20, 21] and allows for good mode confinement to be obtained. Our technique simplifies the fabrication process and enables more complicated dielectric metamaterial architectures (such as vertical stacks of these dielectric metasurfaces).

To fabricate our GaAs metasurfaces we have adapted the selective wet oxidation technique that has been used to form current blocking layers in vertical-cavity surface-emitting lasers.[22] **Figure 1**(a) shows the process flow for creating GaAs resonators starting from a molecular beam epitaxy grown wafer consisting of a semi-insulating GaAs substrate onto which a 300 nm layer of $Al_{0.85}Ga_{0.15}As$ was deposited followed by a 300 nm layer of GaAs. We first deposit a few-hundred nanometers of $SiO_2$ that is used as an etch mask. Next, we spin-coat a positive tone polymethyl methacrylate (PMMA) resist and pattern circular disks using standard electron-beam lithography. After the development of the PMMA, a 10-20 nm layer of nickel is deposited and followed by a lift-off process resulting in thin nickel disks. The shape of the nickel disks is transferred onto the SiO$_2$ layer using inductively-coupled-plasma (ICP) etching. The nickel disks are then removed using nitric acid leaving only the SiO$_2$ disk as an etch mask for GaAs. SiO$_2$ has high etch selectivity (>5) over GaAs/AlGaAs during the Chlorine based ICP etch process. We then use an optimized ICP etch recipe to create GaAs and AlGaAs nanodisks and pillars with smooth vertical side walls. Finally, the sample is placed in a tube furnace at ~420 degree Celsius for selective wet oxidization of the AlGaAs layers. Here, a nitrogen carrier gas is used to transport water vapor across the sample, converting the layers of AlGaAs into their native oxide – low index (Al$_x$Ga$_{1-x}$)$_2$O$_3$.

Note that we can replace the GaAs resonator layers with AlGaAs layers, provided the Al concentration of the resonator layers is at least ~20% lower than that of the oxidation layers. This is because the oxidation rate increases exponentially with Al concentration[20] so that the AlGaAs dielectric resonators will remain mostly unchanged while the under layer is completely oxidized. The realization of AlGaAs dielectric resonators could benefit applications requiring Mie resonances in the visible spectrum since Al$_{0.45}$Ga$_{0.55}$As has a direct bandgap at 624 nm and transitions into an indirect bandgap material when the Al concentration is higher than 45%.[23] Moreover, our fabrication technique can be applied to other Al-containing semiconductors, such as AlInAs, AlInGaP, etc.[20]

Figure 1 (b) & (c) show scanning-electron microscopy (SEM) images of the GaAs nanodisks before and after the oxidation process, respectively. Each nanodisk comprises 3 layers: the topmost dark layer is the residual SiO$_2$ etch mask, and the lower two layers are GaAs and AlGaAs which both appear with a similar light grey color before oxidation due to similar high electronic conductivity. After the oxidation process, the bottom AlGaAs layer appears darker due to the formation of the oxide that has a much lower conductivity. The nanodisks exhibit smooth and slightly tapered vertical sidewalls. The sidewall angle can be optimized by modifying etch parameters such as ICP and reactive ion etch (RIE) power. Since the GaAs resonator heights are fixed by the thicknesses of the epitaxially grown layers, we tuned the wavelengths of the dipole resonances by varying the nanodisk diameters as well as the array pitch while keeping the duty cycle constant at ~40% (we define duty cycle as $d/p$, $d$ being the disk diameter and $p$ the array pitch).

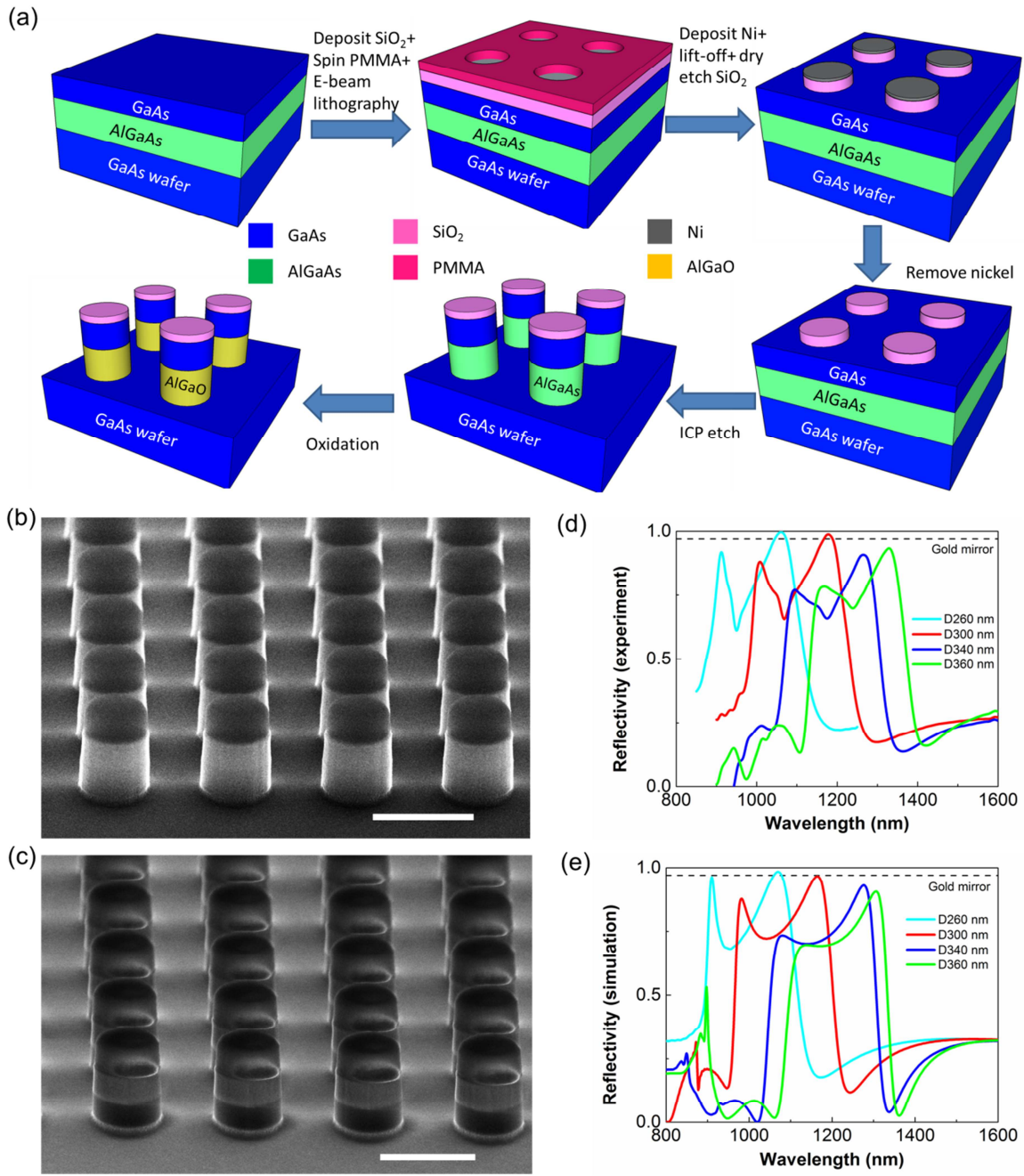

**Figure 1.** (a) Process for fabricating (Al)GaAs based dielectric metasurfaces. The fabrication starts from epitaxial layer growth using MOCVD or MBE, followed by patterning and creating etch masks using e-beam lithography and ICP dry etch, and finished by selectively wet oxidizing AlGaAs to form $(Al_xGa_{1-x})_2O_3$. SEM images at ~75 degree tilted angle showing GaAs dielectric

resonators (b) before and (c) after the selective wet oxidation process. The scale bar corresponds to 1 μm. (d) Experimental and (e) simulated reflectivity spectra of GaAs dielectric metasurfaces with the same thickness of 300 nm and different diameters. The horizontal black dashed line represents ~97% reflectivity from the gold mirror.

We measured the reflectivity of the GaAs dielectric metasurface samples to verify the existence of well separated electric and magnetic dipole resonances which, in turn, would confirm the creation of a fully oxidized spacer layer separating the GaAs nanodisks from the high index substrate. For the measurements, we used a 20X Mitutoyo Plan Apo NIR infinity-corrected objective (numerical aperture = 0.4) to both focus incident white light onto the samples and collect the reflected light. The reflected light was then directed into a near-IR spectrometer and detected using a liquid nitrogen cooled InGaAs camera. The measured reflection spectra were then normalized by the spectrum of a gold mirror measured under the same conditions. Figure 1(d) shows the experimental reflectivity spectra of four different arrays, each with the same resonator height of 300 nm, but with different diameters of 260, 300, 340 and 360 nm. Each spectrum exhibits two reflectivity peaks that correspond to the magnetic and electric dipole resonances, thereby confirming the formation of the native oxide. Moreover, the dipole resonances shift to longer wavelengths as the resonator diameter increases. Finally, the 260 and 300 nm diameter resonators showed near 100% reflection, even outperforming the gold reference mirror at the same wavelengths (gold has a reflectivity of ~97% at ~1 μm shown by the black horizontal dashed line). This near-perfect reflection is attributed to the extremely low residual loss of crystalline GaAs at wavelengths longer than its bandgap (~870 nm).

The experimental data were compared with the results of finite-difference time domain (FDTD) simulations of the electromagnetic response of the resonator arrays. For these simulations we assumed a dispersionless refractive index of 1.6 for the native oxide layer. As seen in Figure 1(e), very good agreement was obtained between the simulated and experimental spectra, further confirming the complete oxidation of the AlGaAs layers and thus the creation of GaAs dielectric metasurfaces. The minor differences between simulation and experiment are attributed to the slight differences in dimensions and shapes between the fabricated and simulated resonators, and that fact that the simulation was performed with plane wave incidence while the experiment was measured with focused light.

As mentioned above, fabricating multilayer or even 3D dielectric metamaterials is possible using techniques presented in this paper. In contrast, fabricating multi-layer Si-based dielectric metamaterials is challenging because: 1) growing multi-layer Si with $SiO_2$ as spacers is not a standard process; and 2) ICP etching of Si and $SiO_2$ requires different recipes that may cause non-uniformities.[7] To demonstrate the ability to produce multilayer metasurfaces, we fabricated 3 stacked layers of the previous GaAs/AlGaO metasurface in a monolithic fashion. The only modification of the single layer process described above is the use of a wafer that has 3 layers of GaAs separated by 3 layers of AlGaAs.

**Figure 2**(a) shows a false-color SEM image of the multilayer sample exhibiting 300 nm thick resonators separated by 300 nm AlGaO native oxide layers. The green, brown and yellow regions correspond to the native oxide layers, the GaAs resonators and the GaAs substrate, respectively. This multilayer structure has a high aspect ratio with a height of >2 μm and a diameter of ~350 nm. Note that this multilayer structure was fabricated with a single step ICP etch since both AlGaAs and GaAs are etched under the same condition. Therefore, we anticipate that 3D dielectric metamaterials are possible since the number of dielectric metasurface layers is in principle limited only by epitaxial growth. In practice, the thickness of etch mask, the strength of semiconductors for supporting high aspect ratio structures, as well as etching conditions will eventually limit the number of layers. Moreover, by adjusting the etching conditions, we can control the taper angle of the multilayer structure so that each GaAs nano-disk layer can have a slightly different diameter leading to different resonant wavelengths for each of the resonator layers.

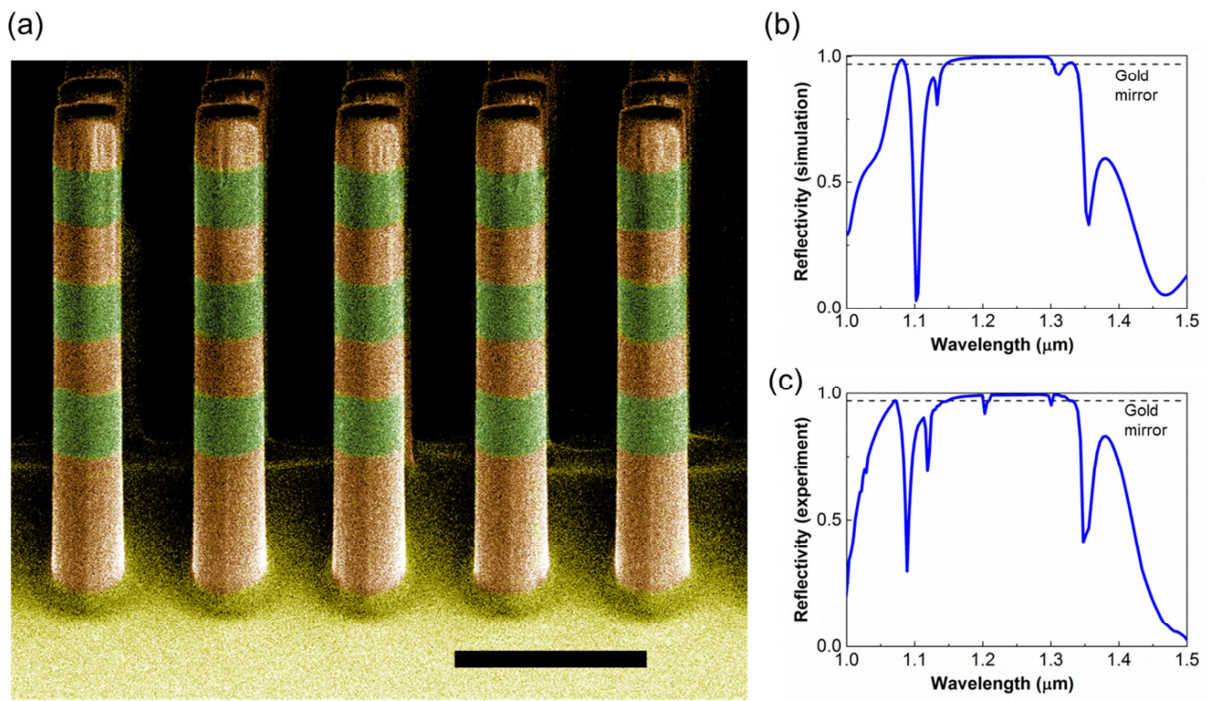

**Figure 2.** (a) A false-color SEM image of a sample comprising three layers of GaAs nano-resonators separated by AlGaO native oxide. Both the GaAs, and the AlGaO layers have heights of ~300 nm. The green, brown, and yellow represent the AlGaO layers, the GaAs resonators and the GaAs substrate, respectively. The nano-post structures are slightly tapered with a diameter

~350 nm at the top and ~370 nm at the bottom. The scale bar corresponds to 1 μm. (b) Simulated and (c) experimental reflectivity spectra of the sample shown in (a) exhibiting ~100% reflectivity over a broad spectral range. The horizontal black dashed line represents ~97% reflectivity from the gold mirror.

The measured reflectivity spectrum of the 3 layer "metasurface" (Figure 2(b)), agrees well with simulation results (Figure 2(c)), and exhibits a higher reflectivity than the gold mirror over a broad (~200 nm) spectral range.[24-27] We note that the spectral range of high reflectivity does not exceed the range between the electric and magnetic dipole resonances of single layer metasurfaces with similar dimensions (comparing to Figure 1(e)) but enhances the reflectivity between the two dipole resonances. Further numerical investigations have shown that by optimizing the diameter and height of the GaAs resonators, unity reflectivity can be achieved over a 400 nm bandwidth at telecom wavelengths (~1.5 μm).

In the following, we show numerical simulations of two different applications that can be enabled by the multilayer all-dielectric metamaterials fabricated using our technique. The first example focuses on independent tuning of the magnetic dipole resonance with respect to the electric-dipole resonance using split nanodisk structures.[28] This approach differs from previous attempts at tuning the relative frequencies of magnetic and electric dipole resonances by changing the aspect ratios of Si nano-disks.[6] **Figure 3**(a) shows a GaAs disk (blue color) separated into two halves by an AlGaO nano-layer (upper yellow color layer). The simulated reflectivity spectra show peaks at longer and shorter wavelengths corresponding to the magnetic and electric dipole resonances, respectively (see supporting information). Figure 3(b) shows that as the gap between the two separated disks increases, the magnetic dipole resonance blue shifts, while the electric dipole resonance remains constant at ~4.4 μm. Eventually the two dipole resonances will overlap (not shown here) which, in turn, achieves the well-studied first Kerker condition.[28]

The second example targets ultrashort pulse recompression for applications such as multiphoton microscopy and lithography.[29] Such recompression typically requires complicated multi-stack dielectric mirrors and/or grating and prisms. Single layer Si-based Huygens' metasurfaces, capable of creating a desired discrete phase gradient, have been successfully employed to demonstrate beam deflection[6],[8] vortex beam conversion[7],[9] 2D holograms,[30] etc. However, it was suggested that multilayer Huygens' metasurfaces might also introduce enough dispersion to recompress femtosecond pulses that have been broadened due to passage through a thick dispersive media.[29] In the following, we show that multilayer Huygens metasurfaces made from III-V semiconductors (Figure 3(c)) are excellent candidates to provide strong group-delay dispersion (GDD). Figure 3(d) shows that both single-layer and three-layer GaAs Huygens metasurfaces transmit over 92% between 900 nm and 1100 nm. The three-layer structure creates a ~6π phase change which is 3 times as large as that of the single layer as shown in Figure 3(e). Furthermore, Figure 3(f) shows that the three-layer structure whose overall thickness is less than 2 μm can provide a group-delay dispersion (GDD) as large as -3000 $fs^2$. Although this simulation was performed with a central wavelength of 960 nm (below the GaAs bandgap) to minimize the absorption loss, we can apply the same technique to 700 nm or even visible wavelengths using AlGaAs (Al>30%) resonators to cover the whole tuning range of Ti:sapphire lasers.

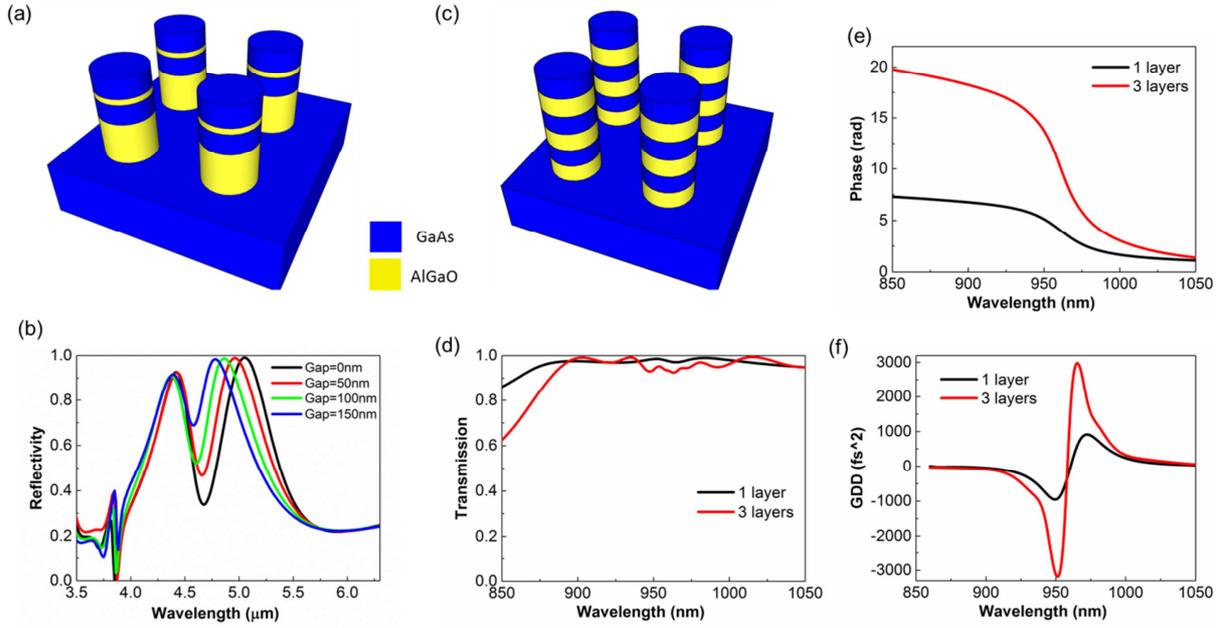

**Figure 3.** (a) Schematic of split GaAs dielectric resonators separated by a thin native oxide layer. (b) Simulated reflectivity spectra of split GaAs nano-resonators (AlGaO thickness of 0, 50, 100 and 150 nm) showing independent tuning of the magnetic dipole resonance. The original un-separated GaAs resonators have a height and diameter both of 1400 nm. (c) Schematic of a multilayer GaAs Huygens' structure that can provide strong GDD. (d) Both the single layer and three layer Huygens' structures exhibit almost unity transmission. (e) Single-layer and three-layer structures provide strong phase response covering 0-2π and 0-6π, respectively. (f) The three-layer Huygens' structure provides >3000 $fs^2$ GDD that is ~3 times the GDD of the single-layer Huygens' surface.

In summary, we have demonstrated the monolithic fabrication of III-V semiconductor-based dielectric metasurfaces and metamaterials using a process that forms a low refractive index native oxide layer between the resonators and the semiconductor substrate. These results are confirmed by the observation of well-defined electric and magnetic dipole resonances in the experimental reflectivity spectra. We further adapted this process to create a multilayer GaAs metamaterial structure whose reflectivity exceeds that of gold over a broad spectral range. We anticipate that further extension of the multilayer strategy could lead to fully 3D metamaterials. We stress that this fabrication technique can be adapted to produce dielectric metamaterials made from other III-V semiconductor materials. Finally, we presented two examples of passive applications that could be enabled by these multilayer structures. We envision that these III-V based all-dielectric metamaterials can also lead to novel active applications such as efficient nonlinear frequency converters, and light emitters, detectors, and modulators.

**Acknowledgements**

Parts of this work were supported by the U.S. Department of Energy, Office of Basic Energy Sciences, Division of Materials Sciences and Engineering and performed, in part, at the Center for Integrated Nanotechnologies, an Office of Science User Facility operated for the U.S. Department of Energy (DOE) Office of Science. Sandia National Laboratories is a multi-program laboratory managed and operated by Sandia Corporation, a wholly owned subsidiary of Lockheed Martin Corporation, for the U.S. Department of Energy's National Nuclear Security Administration under contract DE-AC04-94AL85000.

# III-V semiconductor nano-resonators—a new strategy for passive, active, and nonlinear all-dielectric metamaterials

*Sheng Liu,* [1,2,*] *Gordon A. Keeler*[1]*, John L. Reno*[1,2]*, Michael B. Sinclair*[1]*, Igal Brener*[1,2,*]
[1]Sandia National Laboratories, Albuquerque, NM 87185, USA
[2]Center for Integrated Nanotechnologies, Sandia National Laboratories, Albuquerque, NM 87185, USA
E-mail: snliu@sandia.gov, ibrener@sandia.gov

S1 Electric field distributions and vector plots at the electric and magnetic dipole resonances

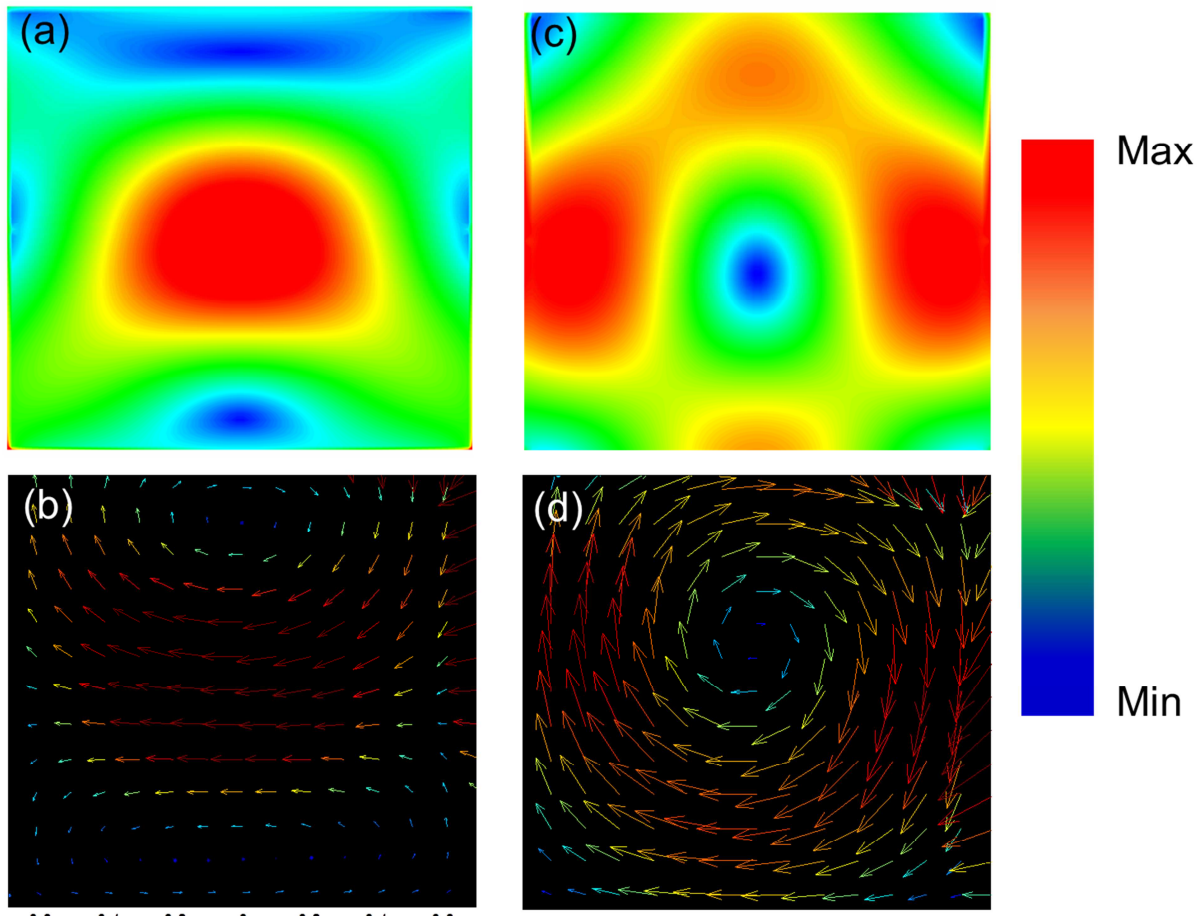

Figure S1. FDTD simulation of the electric field distributions and vector plots in the x-z planoe of a GaAs resonator at wavelengths of 5.1 μm ((a) and (b)) and 4.4 μm ((c) and (d)). These field patterns and vector plots demonstrate magnetic dipole behavior at 5.1 μm, and electric dipole behavior at 4.4 μm.

17